\documentclass[aps,prl,
superscriptaddress,
preprintnumbers,
bibnotes,
amsmath,
amssymb,
twocolumn,
floatfix,
]{revtex4-1}

\usepackage{graphicx}% Include figure files
\usepackage{dcolumn}% Align table columns on decimal point
\usepackage{bm}% bold math
\usepackage[usenames,dvipsnames]{color}
\usepackage{verbatim}
\usepackage{amsfonts}
\usepackage{longtable}
\usepackage{natbib}
\usepackage{hyperref}% add hypertext capabilities
\usepackage[mathlines]{lineno}% Enable numbering of text and display math
\usepackage{dcolumn}
\newcolumntype{d}[1]{D{;}{.}{#1}}

\graphicspath{{figs/}}% Put all figures in a `figs' folder

\def\slashed#1{\kern+0.10em /\kern-0.50em #1}

\begin{document}

\title{Erratum: Standard-model prediction for direct CP violation in $K\to\pi\pi$ decay}

\author{Z.~Bai}%\affiliation{\cu}
\author{T.~Blum}%%\affiliation{\uconn}
\author{P.A.~Boyle}%\affiliation{\edinb}
\author{N.H.~Christ}%\affiliation{\cu}
\author{J.~Frison}%\affiliation{\edinb}
\author{N.~Garron}%\affiliation{\pu}
\author{T.~Izubuchi}%\affiliation{\bnl}%\affiliation{\riken}
\author{C.~Jung}%\affiliation{\bnl}
\author{C.~Kelly}%\affiliation{\riken}
\author{C.~Lehner}%\affiliation{\bnl}
\author{R.D.~Mawhinney}%\affiliation{\cu}
\author{C.T.~Sachrajda}%\affiliation{\soton}
\author{A.~Soni}%\affiliation{\bnl}
\author{D.~Zhang}%\affiliation{\cu}

%\collaboration{RBC and UKQCD Collaborations}
%\noaffiliation

\date{March 8, 2016}

%\preprint{}

%\keywords{CP violation, flavor physics, lattice QCD} %Use showkeys class option if keyword
                              %display desired
\maketitle

While testing new G-parity evolution code to be used to extend the calculation reported in our letter~\cite{Bai:2015nea}, the authors discovered an error in that calculation. We found that when generating the ensemble of gauge configurations on which that calculation was based, the same random numbers were used in the stochastic evaluation of the fermion force for both the $u$ and $d$ flavors.  (A similar duplication occurred for the $s$ and $s'$ flavors.)   As is described below, we have found evidence for this error in quantities computed using this ensemble.  However, these observed effects are so small that we are confident that the physical  results reported in our letter will be affected by this error on a scale two or three orders of magnitude smaller than the statistical errors on those results.  

We emphasize that the strategy presented in our letter to compute the the $\Delta I=1/2$ contribution to $K\to\pi\pi$ decay is not affected by this programming error, an error that is easily corrected by the proper generation of the random numbers needed to evaluate the fermion force.

Because of this error, the same random numbers that were used to evaluate the contribution to a trace at the position $x$ for the up quark were also used at the position $x'$ for the anti-down quark where $x'$ was related to $x$ by a translation by 12 lattice units in the spatial $\hat y$-direction, except for a few points at the boundaries where the relation was more complex.  We have not found a theoretical interpretation for this error which would allow its effect on the results we reported to be estimated.  

However, we have studied its effects empirically.  A direct comparison of the average plaquette found on the G-parity ensemble generated with the error and on an earlier ensemble~\cite{Arthur:2012yc} with no G-parity boundary conditions gives 0.512239(6) and  0.512239(3)(7) respectively.  (The second number is obtained by extrapolating results from Ref.~\cite{Arthur:2012yc} linearly in the light quark mass and the second error on that number is the difference between the linear and a quadratic extrapolation.)   This comparison was made at the beginning of the $K\to\pi\pi$ calculation and added to our confidence that the gauge evolution code was correct.

Of greater interest is a study of the spatial correlations in the gauge field plaquettes.  For a space-time position $x$ and a specific gauge configuration we define $P_{\mu\nu}(x)$ as the conventional real part of one third of the trace of the product of four links that extend in the positive $\mu$ and $\nu$ directions from the point $x$.  We define the covariance of the distribution of the plaquettes with a space-time separation $\delta$ by
\begin{equation}
\mathrm{Cov}(\delta) = \left\langle \frac{1}{6V} \sum_{x,\mu<\nu} \left[
                P_{\mu\nu}(x) P_{\mu\nu}(x+\delta) -\mathcal{P}^2\right] \right\rangle
\end{equation}
where $0 \le \delta_\mu < L_\mu/2$,  $V=L_xL_yL_zL_t$ and for the case at hand $L = (32,32,32,64)$.  The angle brackets indicate an average over gauge configurations and $\mathcal{P}$ is the average plaquette:
\begin{equation}
\mathcal{P} = \left\langle \frac{1}{6V} \sum_{x,\mu<\nu} P_{\mu\nu}(x) \right\rangle.
\end{equation}

\begin{figure}[htb]
\begin{center}
\includegraphics[width=\linewidth]{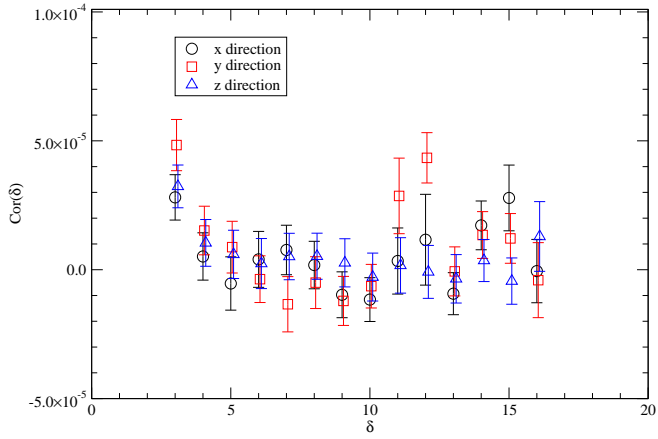}
\end{center} 
\caption{Correlations between the values of the gauge plaquette separated by distances shown on the abscissa for each of the spatial directions.  A very small but statistically significant signal is seen for plaquettes separated in the $\hat y$-direction by 11 and 12 units.} 
\label{fig:corr}
\end{figure}

We define the correlation between spatially separated plaquettes as their normalized covariance:
\begin{equation}
\mathrm{Cor}(\delta) = \frac{\mathrm{Cov}(\delta)}{\mathrm{Cov}(0)}
\end{equation}
The correlation between plaquettes at fixed separation in each of the three spatial directions is shown in Fig.~\ref{fig:corr}.   As can be seen, there is a statistically significant, three standard deviation, correlation in the $\hat y$ direction near the expected separation of 12, a correlation that is not present at other separations or in other directions.  This correlation provides direct evidence that the duplication of the random numbers that occurred in the generation of our gauge ensemble has an observable consequence.  

However, the extremely small size of the correlation seen suggests that the effects of this error in the evolution code are indeed very small.  The error being analyzed occurs in the background gauge ensembles not directly in the measured Green's functions from which our kaon decay  amplitudes are calculated.  Thus, the effects of this error on our results will be similar to the effects seen in the much more sensitive plaquette correlation shown in Fig.~\ref{fig:corr}.  In contrast to these plaquette measurements, the physical results presented in our letter carried fractional statistical errors that were  between one hundred and one thousand times larger than the subtle effects shown in that figure.

We conclude that this error should be expected to influence our published results at a scale that is two or three orders of magnitude smaller than their statistical errors.  However, our lack of theoretical control of the effect of this error and need to rely on empirical methods to bound its effects, detracts from the first-principles basis of our calculation and will be corrected in the future, larger statistics calculation that has been planned.

\bibliography{refs}

\end{document}